# Interactive Traceability Querying and Visualization for Coping With Development Complexity

## Position Paper – Dagstuhl Seminar 12442
## "Requirements Management – Novel Perspectives and Challenges"

Patrick Mäder
Softwaresystems Group
Technische Universität Ilmenau, Germany
patrick.maeder@tu-ilmenau.de

**Motivation**

The engineering of a system can be a complex task. A common way to cope with this complexity is modeling the product to be developed through different levels of abstraction, with different intents, and from different perspectives [OMG03, OMG07]. During software engineering, models are for example used to represent the requirements, the design and the implementation of a software system. As all models of one development project describe different aspects of the same product, they are related by manifold dependencies. For example, the design of a software system depends on its requirements, and the implementation depends on the design [OMG07]. Models are composed of artifacts. An explicit traceability relation between two artifacts captures a dependency between them.

The ability to use existing traceability relations in a development project is called requirements traceability. Traceability is a required component of many software development processes. It is said to provide support for numerous software engineering tasks such as requirements validation, impact analysis, coverage analysis, compliance verification, and derivation analysis. However, despite its potential usefulness, studies have shown that developers and other project stakeholders often create traceability links only because they are required to by external regulations or by process improvement initiatives. Although the required link creation process serves a useful purpose for helping to validate that the system being constructed meets its requirements, studies have shown that stakeholders rarely re-use traceability links during the long-term use and maintenance of a system [RaJa01, ArRi05, Mae+09]. This failure can be partially attributed to the fact that current tools make it difficult for project stakeholders to actually use captured traceability relations. We argue that a more thorough traceability usage support in tools could help to cope with rising complexity in development projects.

**Vision**

Our vision is interactive, context-specific traceability usage. Stakeholders should easily be able to formulate queries that retrieve and employ traces and other development artifacts in order to fulfill a user's information needs. The retrieved information should be visualized in an appropriate way depending on, for example, the information's characteristics, the type of stakeholder, and the currently performed development task. If alternative visualizations exist, the user may choose among them. A visualized static query result is a good starting point and would advance the state of practice already, but we envision the possibility to explore the retrieved results. That means that the user can follow relations of retrieved artifacts and can interactively concretize the initial query.

**Challenges to Address**

*1) Develop user-friendly ways of formulating predefined and ad-hoc traceability queries, involving traces, artifacts, and all their relationships.* – The construction of useful traceability queries is a non-trivial task for users of a software development tool. Trace users are often required to write their trace queries as complex statements using a general query language such as SQL. We approached that problem by proposing the Visual Trace Modeling Language (VTML) [MaCH12]. VTML facilitates predefining traceability queries in a graphical way and to execute these queries when desired. In a user study, we could demonstrate that VTML makes the handling of traceability queries considerably easier for stakeholders. However, our current work did not focus on the integration of VTML into the work environment of users. In addition to predefined queries, users should be able to query artifacts and traces in an ad-hoc way.

*2) Find and develop appropriate visualizations for different types of development artifacts and their traces.* – Traces span graphs across development artifacts. A traceability query returns a subset of those graphs or parts of them. While for some purposes a simple textual representation of results might be suitable, for others a visual representation is in favor. We propose to categorize different results types of traceability queries and to identify appropriate visualizations for them.

*3) Allow browsing of retrieved results and incorporate efficient zoom techniques for abstract and specific views on retrieved data.* – We propose that the result of a query should not be static, but instead further explore-able by the user. Imagine a query that returns for each failed test case the associated requirement. It seems natural that a user of that query may also want to find, for example, other test cases that evaluate the same requirement or related information explaining the requirement (e.g. associated regulatory codes).

*4) Allow interactive filtering of retrieved data and ad-hoc query refinement.* – In addition to allow browsing of returned results, there should also be an option to further filter results and to refine a query within the results.

**Conclusions**

Requirements traceability can in principle support stakeholders coping with rising development complexity. However, studies showed that practitioners rarely use available traceability information after its initial creation. In this position paper for the Dagstuhl seminar 1242, we argued that a more integrated approach allowing interactive traceability queries and context-specific traceability visualizations is needed to let practitioner access and use valuable trace-ability information. The information retrieved via traceability can be very specific to a current task of a stakeholder, abstracting from everything that is not required to solve the task.

**References**

[OMG03]  Object Management Group (OMG), MDA Guide Version 1.0.1, Framingham, Massachusetts, omg/2003-06-01.

[OMG07]  Object Management Group (OMG), Unified Modeling Language Specification (OMG UML) Version 2.4.1, Framingham, Massachusetts. ISO #19505-1:2012, formal/2012-05-06.

[RaJa01] B. Ramesh and M. Jarke, "Toward reference models of requirements traceability", IEEE Transactions on Software Engineering, vol. 27, no. 1, pp. 58–93, 2001.

[ArRi05] P. Arkley and S. Riddle, "Overcoming the traceability benefit problem," in Proceedings 13th International Requirements Engineering Conference, pp. 385–389, IEEE Computer Society, 2005. ISBN 0-7695-2425-7.

[Mae+09] P. Mäder, O. Gotel, and I. Philippow, "Motivation Matters in the Traceability Trenches," in Proceedings of 17th International Requirements Engineering Conference (RE'09), Atlanta, Georgia, USA, pp. 143–148, August 2009.

[MaCH12] P. Mäder and J. Cleland-Huang, "A visual language for modeling and executing traceability queries", Journal of Software and Systems Modeling, DOI: 10.1007/s10270-012-0237-0, 2013.